\documentclass[aip,amsmath,amssymb,reprint,longbibliography]{revtex4-2}

\usepackage[colorlinks,citecolor=blue,linkcolor=red,urlcolor=blue]{hyperref}
\usepackage{color}
\usepackage{graphicx}
\usepackage{subfigure}
\usepackage{verbatim}
\usepackage{fourier}

\usepackage{hyperref}
\usepackage{graphicx}
\usepackage{dcolumn}
\usepackage{bm}
\usepackage{xcolor} 
\usepackage{orcidlink}
\definecolor{color1}{rgb}{0,0.25,0.70}

\begin{document}

	\title{Crystal Orientation Dependence of Extreme Near-Field Heat Transfer between Polar Materials Governed by Surface Phonon Modes}
	
\author{Wei-Zhe Yuan~\orcidlink{0009-0001-4678-4344}}
\affiliation{School of Energy Science and Engineering, Harbin Institute of Technology, Harbin 150001, China}
\affiliation{Key Laboratory of Aerospace Thermophysics, Ministry of Industry and Information Technology, Harbin 150001, China}
\author{Yangyu Guo~\orcidlink{0000-0003-2862-896X}}
\affiliation{School of Energy Science and Engineering, Harbin Institute of Technology, Harbin 150001, China}
\affiliation{Key Laboratory of Aerospace Thermophysics, Ministry of Industry and Information Technology, Harbin 150001, China}
\author{Hong-Liang Yi~\orcidlink{0000-0002-5244-7117}}
\email[]{yihongliang@hit.edu.cn}
\affiliation{School of Energy Science and Engineering, Harbin Institute of Technology, Harbin 150001, China}
\affiliation{Key Laboratory of Aerospace Thermophysics, Ministry of Industry and Information Technology, Harbin 150001, China}
	\date{\today}
	\begin{abstract}
Due to the rapid development of micro- and nano-manufacturing and electronic devices, heat transfer at the transition regime between radiation and conduction becomes increasingly important. Recent work has demonstrated the importance of nonlocal optical response and phonon tunneling.  However, it remains unclear how
the crystal orientation impacts them. In this work, we study this effect on heat transport across vacuum gaps between magnesium oxide (MgO) by nonequilibrium molecular dynamics (NEMD) simulation. At 5~\AA~gaps, the overall thermal conductance exhibits 30\% enhancement for [100] orientation versus [110] and [210], while becoming orientation-insensitive beyond 6~\AA. When the gap size is extremely small, the crystal orientation significantly impacts the resonance frequencies of spectral thermal conductance which are quite close to those of unique surface phonon modes distinct from bulk counterparts. As the gap size gradually increases, the spectral thermal conductance gradually converges to the predicted results of fluctuation-electrodynamics (FE) theory in the long-wavelength approximation. Our findings reveal how surface phonon modes govern
extreme near-field heat transfer across nanogap, providing insights for thermal management in electronic devices.
	\end{abstract}
	
	\maketitle 

Near-field radiation heat transfer across a subwavelength vacuum gap can surpass far-field blackbody limit due to evanescent electromagnetic wave tunneling~\cite{PhysRevB.4.3303}. However, when the separation gap becomes smaller than a few nanometers, corresponding to the transition regime between radiation and conduction, it remains an open question to fully understand the physics drives the heat transport~\cite{ALKURDI2020119963,2014Xiong,2015Chiloyan,Kloppstech2017,Cui2017,Xiong2020}. Advancing fundamental understanding of this regime is critical for scanning thermal microscopy~\cite{2012nl}, heat-assisted magnetic recording~\cite{ieee}, nano-lithography~\cite{2004nl} and near-field radiative energy conversion~\cite{PhysRevApplied.19.037002}.

In the extreme near-field regime, heat transfer mechanisms include near-field radiation~\cite{PhysRevLett.121.045901,PhysRevLett.131.086901}, electron tunneling~\cite{PhysRevB.97.195450,PhysRevB.107.125414}, and, notably, phonon tunneling~\cite{PhysRevLett.94.085901,PhysRevB.106.205418,PhysRevB.105.045410,PhysRevB.106.085403,PhysRevResearch.4.033073,PhysRevB.109.235411,MANCARDOVIOTTI2025116232}. Recently, MD simulation has emerged as an essential tool to uncover the underlying physics in the extreme near-field regime~\cite{CHEN2021121431,D2CP01094A,D3NR00533J,LI2025126945}, and it also serves as a criterion for assessing the validity of continuum fluctuational electrodynamics theory~\cite{PhysRevB.108.L201402,PhysRevB.108.085434,2024APL}. Li \textit{et al.} demonstrate that identical atomic surface terminations in SiC nanogaps enable interfacial thermal resonance~\cite{D3NR00533J} and that the local vibrational density of states (VDOS) at interfacial layers is strongly correlated with spectral thermal conductance. Moreover, Viloria \textit{et al.} identify nonlocal optical response~(i.e. dependence of the
dielectric properties on the wave vector) from both acoustic and optical phonons can influence heat transport in the extreme near-field regime~\cite{PhysRevB.108.L201402}. Guo \textit{et al.} further show deviations between NEMD and continuum FE theory at Ångstrom-nanometer gaps, attributed to phonon tunneling and nonlocal optical response~\cite{PhysRevB.108.085434,2024APL}.​
However, the nonlocal optical response theory relies on macroscopic bulk phonon properties and the homogenization assumption, which neglects the contributions of surface phonon modes and atomic details~\cite{Shan_2024}. The transition of phonons from the bulk to the surface~\cite{liu2025probing,XU2025127295}, as well as its effect on the spectral thermal conductance in the extreme near-field regime, remains poorly understood.

Furthermore, nonlocal optical response also depends on the
direction of wave propagation in the medium, which can introduce optical anisotropy even in materials with isotropic macroscopic properties~\cite{2014apl,PhysRevB.109.035201}. Consequently, crystal orientation might critically modulate extreme near-field heat transfer in polar materials through these anisotropic optical responses and atomic details. Although there have been several studies on the influence of crystal orientation and anisotropy on interface thermal transport~\cite{QI2025109231,PhysRevB.84.125408,2012JAP,MONACHON20138,WEI2022103147,Wang22012024}, their impact remains unexplored in the extreme near-field regime.

In this work, by employing NEMD simulations, we study the effect of crystal orientation on extreme-near field heat transport between two parallel MgO plates (Fig.~\ref{fig:str}, see Sec.~\uppercase\expandafter{\romannumeral1} of Supplemental Material~\cite{sup} for details\nocite{LAMMPS,HIREL2015212,PhysRevB.108.085434,BKS,MgO1989Born,PhysRevB.111.184310,EASTWOOD1980215,PhysRevB.90.134312,PhysRevB.91.115426,PhysRevE.93.052141,PhysRevB.106.085403,PhysRevB.108.085434,PhysRev.123.777,Gangemi_2015,2014YangJY,Song2015AIP,PhysRev.188.1407}).  As the gap size decreases, the influence of atomic details and crystal orientation become increasingly significant, which results in an increasing difference in overall thermal conductance. Furthermore, we found that the surface phonon modes dominate extreme near-field heat transfer by comparing the spectral thermal conductance and VDOS. Besides, we compare the direct NEMD simulation and FE theory with local dielectric function calculated
from equilibrium molecular dynamics (EMD). As the gap size increases, spectral thermal conductance will gradually converge and become independent of crystal orientation and surface phonon modes.

\begin{figure}[htbp]
	\centering
	\includegraphics[width=1\linewidth]{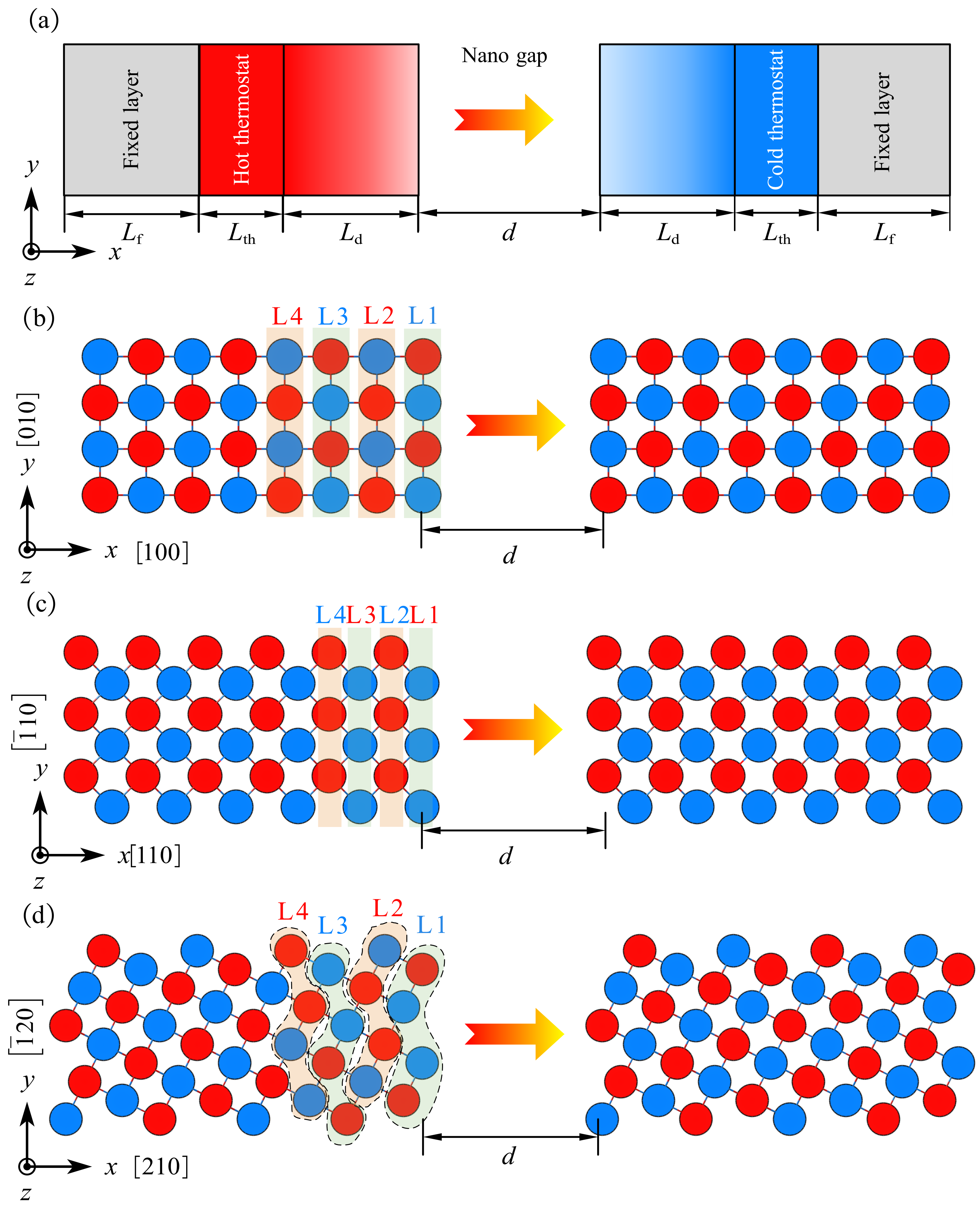}
	\caption{Schematic of NEMD model of MgO-MgO nanogap with gap size $d$: (a) Lengths of fixed layer and thermostat are $L_\mathrm{f}$ and $L_\mathrm{th}$, respectively, whereas length of device region on one side of nanogap is $L_\mathrm{d}$. Schematic of atomistic model with crystal orientation: (b) [100], (c) [110] and (d) [210]. Blue and red atoms denote Mg$^{2+}$ and O$^{2-}$ ions, respectively. The shadow area indicate the zones where the local vibration density of state are extracted (labeled as L1, L2, L3,
		and L4). The gap size is the distance between the center of surface atoms.}
	\label{fig:str}
\end{figure}

\begin{figure}[htbp]
	\centering
	\includegraphics[width=0.7\linewidth]{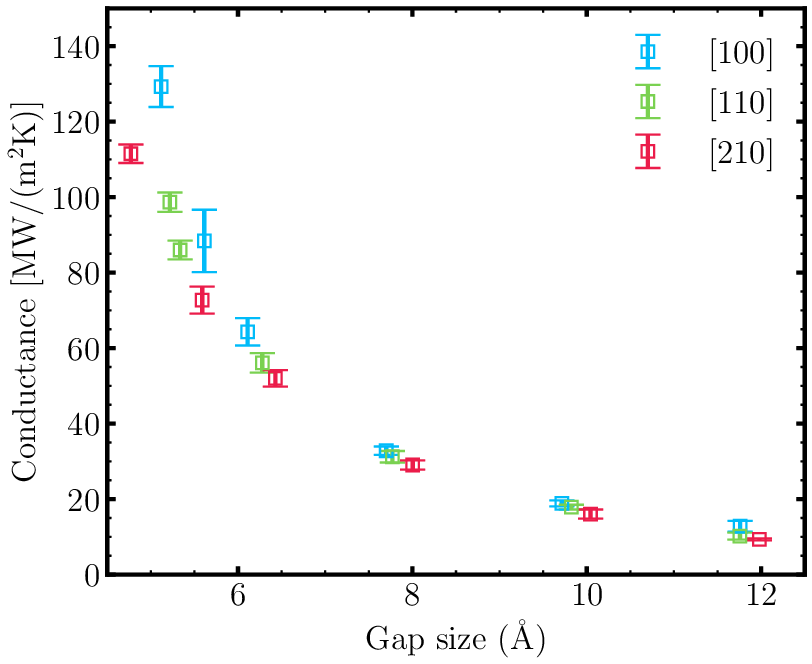}
	\caption{Thermal conductance of MgO-MgO nanogaps vs gap size $d$ at 300 K with different crystal orientations by NEMD. The blue, green and red squares
		with error bars denote the present results with different crystal orientations.}
	\label{fig:Gd1}
\end{figure}

The overall quantum corrected thermal conductances of MgO-MgO nanogaps with different crystal orientations at 300 K are shown in Fig.~\ref{fig:Gd1}.
Due to differences in crystal orientation, the gap size after structure relaxation will differ slightly from the initial given value, especially at small gap sizes. It is worthy of noting that the MgO-MgO nanogaps cannot be stable at a gap size smaller than 4.5 \AA~due to the strong short-range attraction. In the range of gap sizes from 6 \AA~to
12 \AA, the overall thermal conductance across the gap by NEMD is nearly the same and independent of crystal orientation. However, a considerable difference in overall thermal conductance emerged when the gap size was 5 \AA. At this gap size, overall thermal conductance is nearly equivalent for the [110] and [210] crystal orientations. In contrast, the [100] crystal orientation exhibits approximately 30\% greater conductance than the former two orientations.

\begin{figure*}[htbp]
	\centering
	\includegraphics[width=1\linewidth]{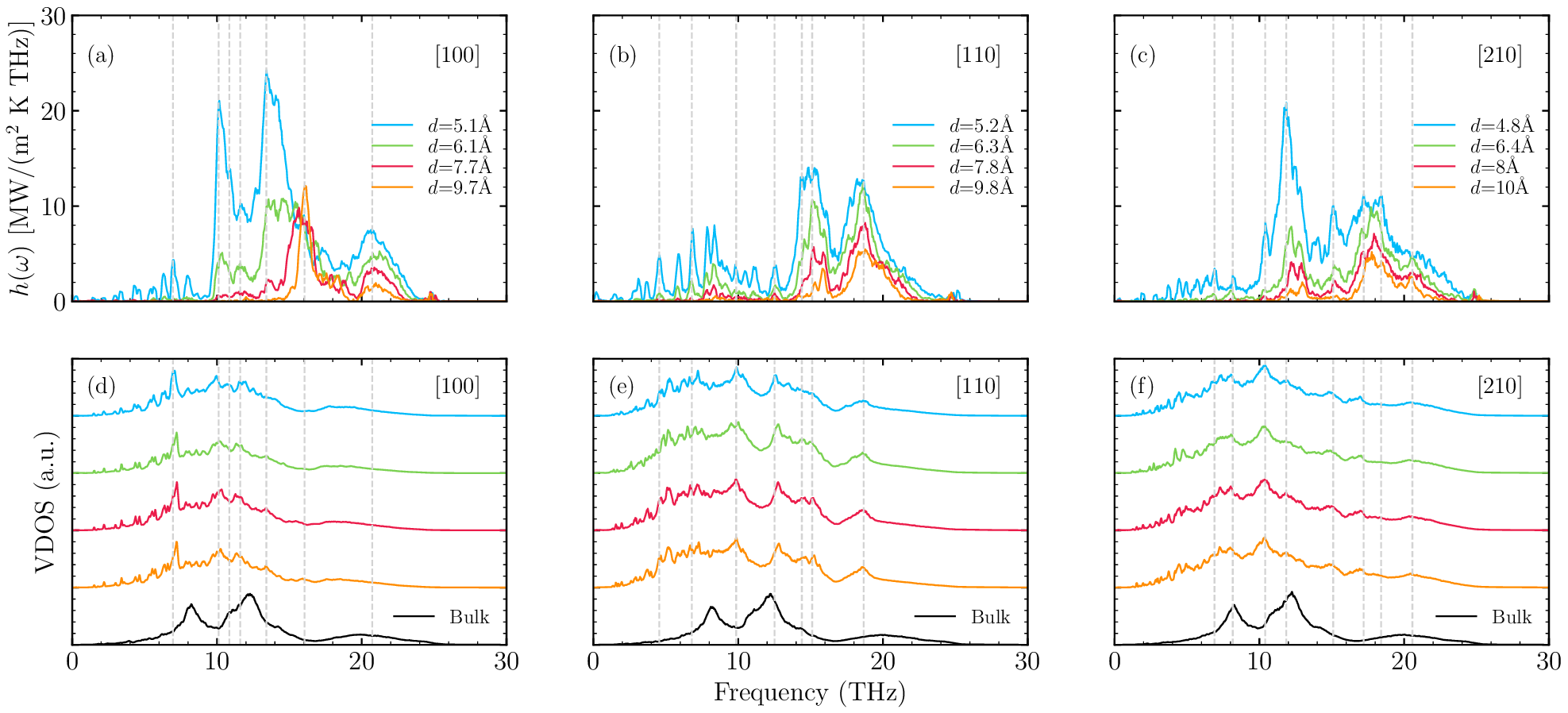}
	\caption{Spectral thermal conductance across MgO-MgO nanogap at various gap sizes with crystal orientations: (a) [100], (b) [110] and (c) [210]. (d)-(f) The normalized local VDOS of surface layer in MgO-MgO nanogap at various gap sizes with different crystal orientations. Grey dashed
		lines are used for guiding the eye.}
	\label{fig:shc}
\end{figure*}

To provide a deeper understanding, we analyze the spectral thermal conductance of the MgO-MgO nanogap with the different crystal orientations. The spectral thermal conductance generally decreases in both spectral range and magnitude as the gap size increases as shown in Fig.~\ref{fig:shc} (a-c). In addition, the alteration in the crystal orientation exerts a substantial influence on the spectral distribution of thermal conductance. 

\begin{figure*}[htbp]
	\centering
	\includegraphics[width=0.75\linewidth]{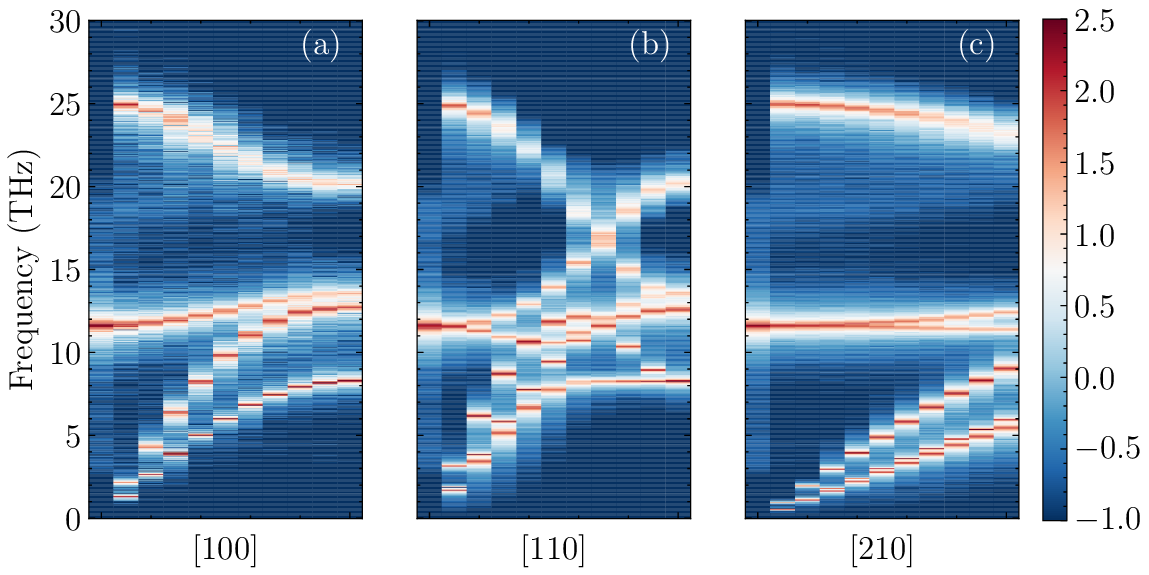}
	\caption{Imaginary part of nonlocal dielectric function at 300 K in log scale (i.e., log $\varepsilon''$) of bulk MgO showing phonon dispersion along: (a) [100], (b) [110] and (c) [210] directions.}
	\label{fig:epsnl}
\end{figure*}

For a gap size of approximately 5 \AA, energy transmission occurs in the whole frequency range corresponding to both acoustic and optical phonons. For these three crystal orientations, the spectral thermal conductance displays small peaks within the range of $0\sim10$ THz, attributed to the longitudinal acoustic~(LA) and transverse acoustic~(TA) phonons based on the phonon dispersion (Fig.~\ref{fig:epsnl}, extracted from nonlocal dielectric function~\cite{2014apl,PhysRevB.108.085434}).  In the range of $10\sim15$ THz,  the crystal orientation exerts a significant influence on the spectral thermal conductance peak ([100]: 10.11 and 13.41 THz; [110]: 15 THz; [210]: 11.86 THz). For the [100] crystal direction, phonons exhibit strong tunneling in this frequency range, resulting in an increase in the overall thermal conductance compared to the other two crystal directions.
In the range of $15\sim25$ THz, crystal orientation also has a significant impact on the spectral thermal conductance. The heat transport in this spectrum may be contributed from the surface
phonon polaritons~(SPhP) formed by the hybridization of transverse and longitudinal optical (TO and LO) phonons with electromagnetic waves as discussed in previous work~\cite{PhysRevB.108.085434,PhysRevB.108.L201402,2024APL}. Based on the resonance frequency of SPhP ($\mathrm{Re}\,\varepsilon = -1$), the spectral thermal conductance based on nonlocal FE theory should have a blue shift compared to the TO and LO phonon frequencies. However, the spectral thermal conductance peak is not observed above 20 THz for crystal orientation [210], which is inconsistent with the prediction of the LO non-local dielectric function defined in these works~\cite{PhysRevB.108.085434,PhysRevB.108.L201402,2024APL}. Actually, this nonlocal FE theory relies on macroscopic bulk phonon properties and homogenization assumption. This phenomenon demonstrates that extreme near-field heat transport cannot be explained solely by bulk phonon properties, suggesting
the presence of additional phonon contributions originating from the surface.

To reveal the effect of surface phonon
modes, Fig.~\ref{fig:shc} (d-f) analyze the local VDOS of surface layer (L1 group in Fig.~\ref{fig:str}) in MgO-MgO nanogap with different crystal orientations. The surface layer exhibits
a distinctive spectral signature compared to the bulk
material, and it is highly sensitive to the arrangement of atoms at the surface. Moving to the L2 group of MgO, the influence of surface atomic morphology on the vibration properties still exists as shown in Fig.~\textcolor{blue}{S1}.
Starting from the L3 group, the VDOS gradually stabilizes and becomes indistinguishable from layers farther from the nanogap, indicating a transition to bulk phonon characteristics. The layer-resolved VDOS analysis clearly shows that the surface phonon modes differ substantially from those in the bulk regions. These pronounced differences arise from the breaking of translational symmetry and the distinct atomic environment near the gap, which leads to the emergence of these novel phonon modes at the surface.
Furthermore, we compare the spectral thermal conductance and surface VDOS in Fig.~\ref{fig:shc}. Most peaks in the spectral thermal conductance align with the surface VDOS, indicating that these surface
modes provide a dominant pathway for phonon tunneling across the MgO-MgO nanogap, that
are fundamentally distinct from those in the bulk material. Nevertheless, there are some peaks of spectral thermal conductance that are not matched with the surface VDOS, which is probably due to the contribution of SPhP ([100]: 16.05 and 20.72 THz; [210]: 18.41 THz). No such mismatch was observed in the [110] crystal orientation, which might be related to the non-polar arrangement of atoms as shown in Fig.~\ref{fig:str} (c). Although our results highlight the correlation between surface phonon modes and spectral thermal conductance, a comprehensive physical understanding of their roles in thermal transport remains a significant challenge.

\begin{figure}[htbp]
	\centering
	\includegraphics[width=0.8\linewidth]{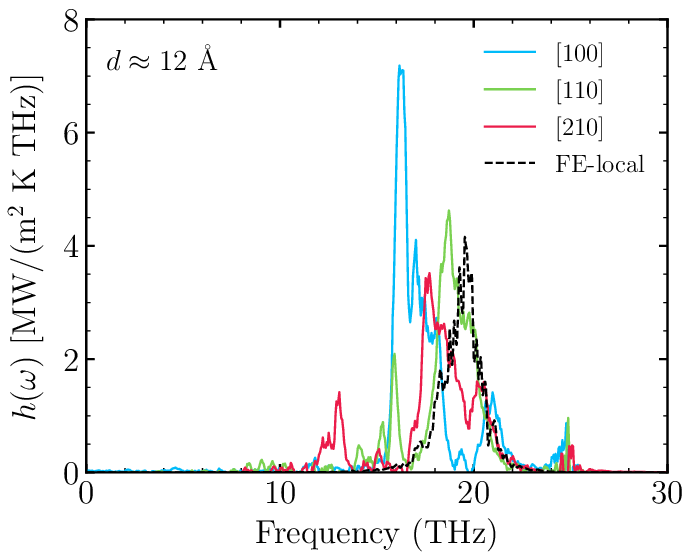}
	\caption{Comparison of spectral thermal conductance of MgO-MgO nanogaps at gap size of 12~\AA~obtained from NEMD with different crystal orientations and FE local theory.}
	\label{fig:shc12}
\end{figure}

As the gap size increases, there is no significant change in the surface VDOS, while the spectral thermal conductance contributed by acoustic phonons almost disappears as shown in Fig.~\ref{fig:shc}. Although the non-local FE prediction shows that the spectral thermal conductance will undergo a frequency shift as the gap size increases since the excitation wave vector tends toward the $\Gamma$ point~\cite{PhysRevB.108.L201402}, this phenomenon has not been observed in NEMD. The physical picture of the nonlocal optical response of polar materials, as well as the incorporation of surface effects into this framework, remain open questions. Therefore, we only compare the direct NEMD simulation and FE local theory at larger gap size by supplementing the latter with local dielectric function calculated
from EMD (see Sec.~\uppercase\expandafter{\romannumeral2} of Supplemental Material~\cite{sup} for details). As the gap size increases to 12 \AA, the spectral thermal conductance gradually approaches to the local FE theory prediction as shown in Fig.~\ref{fig:shc12}. This phenomenon can be attributed to the predominance of long-wave optical phonons at larger gap sizes~\cite{2015Chiloyan,PhysRevB.108.085434}. The isotropic macroscopic properties of MgO result in a negligible effect of the crystal orientation on heat transfer. Although there are still slight differences in the spectral thermal conductance for different crystal directions at this gap size, possibly attributable to surface effects, we expect spectral thermal conductance will converge to FE local theory at larger nanogaps. But we haven’t simulated it due to significant fluctuations of NEMD at such sizes~\cite{PhysRevB.108.085434}.

Finally, the surface phonon modes which result primarily from symmetry breaking near the gap, are expected to persist over temperatures, unlimited to 300 K.  Thus, the effect of surface phonon modes and crystal orientation on extreme near-field thermal transport is likely to be observable at other temperatures as well.

In summary, this study elucidates the profound impact of crystal orientation on extreme near-field heat transfer between polar materials. Through NEMD simulations, we reveal that the [100] crystal orientation shows 30\% greater overall thermal conductance compared to [110] and [210] at 5~\AA~gaps, whereas no orientation dependence is observed beyond 6~\AA. Furthermore, it was determined that
atomic details and surface phonon modes significantly impact resonance frequencies of phonon tunneling and optical responses at ultra tiny gap size. As the gap size increases, spectral thermal conductance gradually converges and becomes independent of crystal orientation, a phenomenon that aligns with the macroscopic FE theory. 
The present work provides further insight into the physics of heat transport in the extreme near-field regime between polar materials, offering further exploration and application of these phenomena in thermal management in atomic scale.
	
	\hspace*{\fill}
\begin{acknowledgments}	
This paper was supported by the National Natural Science Foundation of China (Grants No. U22A20210 and 52576070), the National Natural Science Foundation for Excellent Young Scientists Fund Program (Overseas) and starting-up funding from Harbin Institute of Technology~(Grant No. AUGA2160500923).
\end{acknowledgments}

		\section*{Data Availability Statement}
		The data that support the findings of this study are available from
		the corresponding authors upon reasonable request.


%

\clearpage
\onecolumngrid
\begin{center}
	\textbf{\large Supplemental Materials for ``Crystal Orientation Dependence of Extreme Near-Field Heat Transfer between Polar Materials Governed by Surface Phonon Modes''}\\[.2cm]
	Wei-Zhe Yuan~\orcidlink{0009-0001-4678-4344}$^{1,2}$, Yangyu Guo~\orcidlink{0000-0003-2862-896X}$^{1,2}$ and  Hong-Liang Yi~\orcidlink{0000-0002-5244-7117}$^{1,2,\textcolor{red}{\left.\text{a}\right)}}$ \\[.1cm]
	{\itshape ${}^1$ School of Energy Science and Engineering, Harbin Institute of Technology, Harbin 150001, China}\\[.1cm]
	{\itshape ${}^2$ Key Laboratory of Aerospace Thermophysics, Ministry of Industry and Information Technology, Harbin 150001, China}\\[.1cm]
	(Dated: \today)
\end{center}

\maketitle
\setcounter{equation}{0}
\setcounter{section}{0}
\setcounter{figure}{0}
\setcounter{table}{0}
\setcounter{page}{1}
\renewcommand{\theequation}{S\arabic{equation}}
\renewcommand{\thesection}{ \Roman{section}}

\renewcommand{\thefigure}{S\arabic{figure}}
\renewcommand{\thetable}{S\arabic{table}}
\renewcommand{\tablename}{Table}

\renewcommand{\bibnumfmt}[1]{[S#1]}
\renewcommand{\citenumfont}[1]{#1}
\makeatletter

\maketitle

\maketitle

\section{NEMD simulation and analysis}
In this work, NEMD simulations of heat transfer through two parallel vacuum-separated MgO plates are performed using the open-source package \textsc{LAMMPS}~\cite{LAMMPS}, with the system geometry detailed in Fig. \textcolor{blue}{1}. MgO crystals (space group: $\mathrm{Fm}\bar{3}\mathrm{m} \left[ 225 \right]$) along the [100], [110] and [210] direction are investigated in this work. In the orthogonal simulation box, MgO with these crystal orientations was generated using Atomsk tool~\cite{HIREL2015212}.  We only study cases with Mg-O atomic surface terminations because the large force between identical atomic layers makes the system unstable. In the NEMD simulation, the nanogap is positioned between hot and cold thermostats with two fixed layers on either end. These fixed layers do not engage in the evolution of atomic dynamics during the steady-state run. They are used to prevent macroscopic drift and ensure stability. Periodic boundary conditions are enforced in all three directions of the system. The length of the fixed-layer region ($L_\mathrm{f}$), thermostat ($L_\mathrm{th}$) and device region ($L_\mathrm{d}$) for MgO nanogap model with different crystal orientation are shown in Table~\ref{tab2}, which are tested to be sufficiently thick to avoid fictitious periodic images \cite{PhysRevB.108.085434}. In addition, the cross sections for the cases with similar or dissimilar crystal orientation are setted in Table~\ref{tab3} to avoid lattice mismatch, which have been verified for size independence. We use a pairwise van Beest, Kramer, and van Santen (BKS) potential form $\phi_{ij}$ between atoms $i$ and $j$ composed of a short-range Buckingham term and a long-range Coulomb interaction which can be written as~\cite{BKS}
\begin{equation}\label{eq:bks}
	\phi _{ij}=\frac{q_iq_j}{r_{ij}}+A\exp \left( -\frac{r_{ij}}{\rho} \right) -\frac{C}{r_{ij}^{6}},
\end{equation}
where $r_{ij}$ is the distance between atom $i$ and atom $j$, $q_i$ and $q_j$ are the partial charges of the atoms. The other atomic parameters for MgO are given in Table~\ref{tab1}~\cite{MgO1989Born}. This potential is selected because it accurately reproduces the dielectric function of bulk MgO in the infrared spectrum \cite{PhysRevB.111.184310}, which is vital to describing near-field heat transport across a vacuum gap. The cutoff radius of the short range interaction in real space is 10~\AA. Additionally, the particle-particle particle-mesh (PPPM) method~\cite{EASTWOOD1980215} is implemented to treat long-range Coulomb forces outside the cutoff radius with an accuracy of $10^{-4}$ in \textbf{k}-space. After the structural optimization, the lattice constant of MgO at 300 K is 4.217~\AA. The nanogap is obtained by shifting half part of a bulk MgO crystal along the $x$ direction.  For the NEMD simulation, a time step of 0.5 fs is used. First, $1\times10^6$ time steps are run to relax the whole
system under the $NPT$ ensemble. Due to the different stress conditions on the bulk materials and the nanogap, there will be an unstable period at the beginning of the MD relaxation process, but it will eventually stabilize. Even though the initial atomic velocity distributions are different, the uncertainty of the gap size after stabilization is less than 0.1 \AA~for our cases at 300 K. Then, the fixed-layer regions are fixed, and $2\times10^6$ time steps are run to allow the remaining free part to reach a steady state under the influence of Langevin thermostats in the $NVE$ ensemble. Finally, $2\times10^6$ time steps of steady-state runs are performed to calculate the spectral and overall thermal conductance of the nanogap. The force and velocity of the system is output once per 20 time
steps during the steady-state run. The temperature differences applied on the two thermostats is $T\pm50$ K when $T=300$ K. To reduce statistical fluctuations, five independent NEMD simulations are conducted for each gap size.

\begin{table}[htbp]
	\centering
	\caption{Length of the fixed-layer region, thermostat and device region of NEMD model of MgO-MgO nanogap.}\label{tab2}
	\begin{ruledtabular}	
		\begin{tabular}{ccccc}	
			gap size (nm)&Orientation &$L_\mathrm{f}$ (nm)& $L_\mathrm{th}$ (nm) & $L_\mathrm{d}$ (nm)   \\  \hline
			&	$[100]$  &3.37& 2.11 &3.37  \\ 
			<1&	$[110]$  &3.58&1.79 & 3.58 \\ 
			&	$[210]$  &3.76 &1.88 & 3.76  \\	 \hline
			&	$[100]$  &5.05& 2.11 &5.05  \\ 
			$\geq1$&	$[110]$  &5.37&1.79 & 5.37 \\ 
			&	$[210]$  &5.64 &1.88 &5.64  \\	 
		\end{tabular}
	\end{ruledtabular}
\end{table}
\begin{table}[htbp]
	\centering
	\caption{Cross section of NEMD model of MgO-MgO nanogap with different crystal orientations.}\label{tab3}	
	\begin{ruledtabular}
		\begin{tabular}{ccc}	
			Orientation &$L_y$ (nm)& $L_z$ (nm)  \\  \hline
			$[100]$-$[100]$  &3.37& 3.37   \\ 
			$[110]$-$[110]$  &3.58&3.37 \\ 
			$[210]$-$[210]$  &3.76&3.37   \\	
		\end{tabular}
	\end{ruledtabular}
\end{table}
\begin{table}[htbp]
	\centering
	\caption{Parameters in BKS potential of MgO~\cite{MgO1989Born}.}\label{tab1}
	\begin{ruledtabular}	
		\begin{tabular}{ccccc}	
			
			Pair of atoms &$q$ (e)& $A$ (eV) & $\rho$ (\AA) & $C$ (eV$\cdot$\AA$^6$) \\  \hline
			Mg-Mg  &$+1.4$& 1309363.25 & 0.104 & 0 \\ 
			O-O  &$-1.4$&2145.86142 & 0.300 & 30.22 \\ 
			Mg-O  & &9892.35947 & 0.202 & 0 \\	 
		\end{tabular}
	\end{ruledtabular}
\end{table}

The entire heat flow from one side ($I$) of the nanogap device region to the opposite side ($J$) can be divided into its spectral component as
\begin{equation}
	Q_{I\rightarrow J}=\int_{0}^{\infty}q(\omega)\frac{d\omega}{2\pi},
\end{equation}
with $\omega$ the angular frequency and the spectral heat flux estimated using the Fourier transform of the time ($t$) correlation function between atomic force and velocity \cite{PhysRevB.90.134312,PhysRevB.91.115426,PhysRevE.93.052141}:
\begin{equation}\label{Eq:qw}
	q\left(\omega\right)=2\mathrm{Re}\sum_{\substack{i=I\\  
			j=J}}\int_{-\infty}^{\infty}\left\langle\mathbf{F}_{j}\left(t\right)\cdot\mathbf{v}_{j}\left(0\right)\right\rangle\exp(i\omega t)dt,
\end{equation}
where $\mathbf{F}_j\left( t \right) =\sum_{i\in I}{\mathbf{F}_{ji}}\left( t \right) $ is the overall force on atom $j$ in region $J$ from all the atoms in region $I$. $\mathbf{v}_{j}$ is the atomic velocity, and the bracket $\left<  \right> $ represents the nonequilibrium ensemble average as calculated by time average.

Using the spectral heat flow computed from the NEMD simulation using Eq.~\ref{Eq:qw}, the transmission function across the nanogap is calculated as \cite{PhysRevB.90.134312}
\begin{equation}
	\Xi\left(\omega\right)=\frac{q\left(\omega\right)}{k_{\mathrm{B}}\Delta T},
\end{equation}
where $k_{\mathrm{B}}$ is the Boltzmann constant, and $\Delta T$ is the temperature difference between the hot and cold thermostats.
According to Landauer's formula, the classical and quantum thermal conductances of the nanogap per unit area can be computed, respectively, as follows \cite{PhysRevB.106.085403}
\begin{equation}\label{eq:cl}
	h_{\mathrm{classical}}=\frac{1}{A_c}\int_0^{\infty}{k_{\mathrm{B}}\Xi (\omega )\frac{d\omega}{2\pi}},
\end{equation}
\begin{equation}\label{eq:q}
	h_{\mathrm{quantum}}=\frac{1}{A_c}\int_0^{\infty}{\hbar \omega \frac{\partial f_{\mathrm{BE}}\left( \omega \right)}{\partial T}\Xi (\omega )\frac{d\omega}{2\pi}},
\end{equation}
where $A_c$ is the cross-section area of device. The classical ($k_{\mathrm{B}}$) and quantum heat capacity $[\hbar \omega \partial f_{\mathrm{BE}}\left( \omega \right)/\partial T]$ are used, respectively, with $\hbar$ the reduced Planck constant, $f_{\mathrm{BE}}\left( \omega \right)$ the Bose-Einstein equilibrium distribution. Unless otherwise specified, all subsequent results of this study are quantum-corrected.
\section{Fluctuational electrodynamics theory with EMD inputs}
Note that the electronic polarization are not taken into account in the MD simulation~\cite{PhysRevB.111.184310}. For a fair comparison,
we calculated the local dielectric function without electronic degrees of freedom as $\varepsilon_{\alpha \beta}(\omega )=\delta _{\alpha \beta}+\chi _{\alpha \beta}\left( \omega \right)$. The details of the EMD simulation can be found in our previous work~\cite{PhysRevB.108.085434,PhysRevB.111.184310}. $\delta_{\alpha\beta}$ is the Kronecker delta, and the imaginary part of susceptibility $\chi''_{\alpha \beta}$ is calculated via the fluctuation-dissipation theorem (i.e. Green-Kubo formula)~\cite{PhysRev.123.777,Gangemi_2015}: 
\begin{equation}\label{eq:chi}
	\chi''_{\alpha \beta}\left( \omega \right)=\mathrm{Im}\frac{1}{\varepsilon _0Vk_{\mathrm{B}}T\omega ^2}\bigg[ \left< J_{\alpha}\left( 0 \right) \cdot J_{\beta}\left( 0 \right) \right> 
	+i\omega \int_0^{\infty}{\left< J_{\alpha}\left( 0 \right) \cdot J_{\beta}\left( t \right) \right> e^{i\omega t}dt} \bigg],
\end{equation}
where $V$ is the system volume and $\varepsilon_0$ represents the dielectric permittivity of the vacuum. In Eq.~(\ref{eq:chi}), the ionic polarization of the system is calculated as the current density $\mathbf{J}\left( t \right)$:
\begin{equation}
	\mathbf{J}\left( t \right) =\sum_{lb}{q_b\mathbf{v}_{lb}\left( t \right)},
\end{equation}
where $\mathbf{v}_{lb}$ is the atomic velocity and $q_{b}$ is the charge of the $b$-th basis atom. The real part of susceptibility $\chi'_{\alpha \beta}$ is then calculated using Kramer-Kronig relation~\cite{2014YangJY}
\begin{equation}
	\chi '_{\alpha \beta}\left( \omega \right) =\frac{2}{\pi}\int_0^{\infty}{\frac{\omega '\chi ''_{\alpha \beta}\left( \omega ' \right)}{\omega '^2-\omega ^2}d\omega '}.
\end{equation}

In the framework of FE, the spectral radiative thermal conductance between two parallel plates separated by a vacuum gap $d$ can be written in a Landauer-like form as~\cite{Song2015AIP}:
\begin{equation}
	h_{\mathrm{FE}}\left( \omega \right) =\int_0^{\infty}{\frac{\partial \Theta \left( \omega ,T \right)}{\partial T}\frac{kdk}{4\pi ^2}\sum_{i=\mathrm{s},\mathrm{p}}{\xi _i\left( \omega ,k,\mathrm{d} \right)},}
\end{equation}
where $\Theta \left( \omega ,T \right) =\hbar \omega /\left[ \exp \left( \hbar \omega /k_BT \right) -1 \right] $ is the mean energy of a Planck oscillator, $k$ is the parallel
wave vector, and $\xi_\mathrm{s}$ and $\xi_\mathrm{p}$ are the photonic transmission coefficient of the $\mathrm{s}$- and $\mathrm{p}$-polarized modes~\cite{Song2015AIP}, respectively:
\begin{equation}\label{eq:7}
	\xi _{i=\mathrm{s, p}}\left( \omega ,k \right) =\begin{cases}
		\frac{\left( 1-\left| r_{i}^{1} \right|^2 \right) \left( 1-\left| r_{i}^{2} \right|^2 \right)}{\left| 1-r_{i}^{1}r_{i}^{2}e^{2i\beta _0d} \right|^2},k<k_0,\\
		\frac{4\mathrm{Im}\left( r_{i}^{1} \right) \mathrm{Im}\left( r_{i}^{2} \right) e^{-2\mathrm{Im}\left( \beta _0 \right) d}}{\left| 1-r_{i}^{1}r_{i}^{2}e^{2i\beta _0d} \right|^2},k>k_0.\\
	\end{cases}
\end{equation}
for propagating ($k<k_0$) and evanescent ($k>k_0$) waves, where $k_0=\omega/c$ is the wave vector in vacuum. $\beta_0=\sqrt{k_{0}^{2}-k^2}$ is the wave vector perpendicular to the surface. $r_{\alpha}^{j}$ is the reflection coefficient for $\alpha$-polarization between vacuum and medium $j$ ($j=1$ for emitter and $j=2$ for receiver) can be written as follows:
\begin{subequations}
	\label{eq:whole}
	\begin{equation}
		r_{\mathrm{s}}^{j}=\frac{\beta _0-\beta _j}{\beta _0+\beta _j},
	\end{equation}
	\begin{equation}
		r_{\mathrm{p}}^{j}=\frac{\varepsilon _j\beta _0-\beta _j}{\varepsilon _j\beta _0+\beta _j},
	\end{equation}
\end{subequations}
where $\beta _j=\sqrt{\varepsilon _jk_{0}^{2}-k^2}$ is the perpendicular wave vector in medium $j$, $\varepsilon _j$ being the dielectric function of medium $j$.

\section{Transition of local vibrational density of states from bulk to surface}
The local vibrational density of states~(VDOS) $\rho(\omega)$  is obtained by performing a Fourier transform of the velocity auto-correlation function~\cite{PhysRev.188.1407}:
\begin{equation}
	\rho \left( \omega \right) =\int_{-\infty}^{\infty}{\sum_j{\mathbf{v}_j\left( 0 \right) \cdot \mathbf{v}_j\left( t \right) e^{i\omega t}dt}}
\end{equation}
\begin{figure}[htbp]
	\centering
	\includegraphics[width=0.85\linewidth]{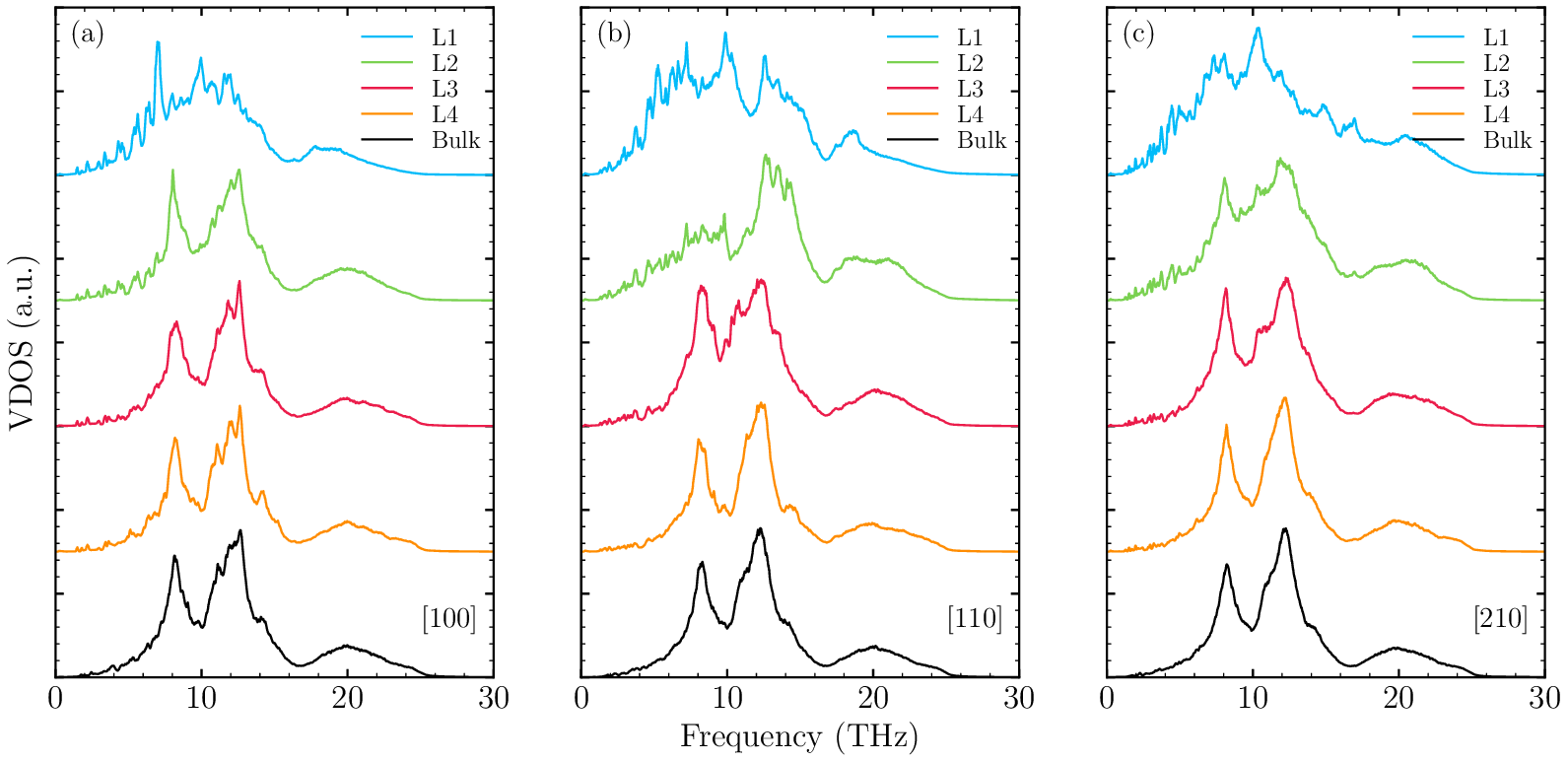}
	\caption{Transition of normalized local vibrational density of states from bulk to surface with different crystal orientations: (a) [100], (b) [110] and (c) [210].}
	\label{fig:vdosl}
\end{figure}

To calculate the
local VDOS profile (see Fig.~\textcolor{blue}{1}), we divide the system into different groups along thermal transport direction, with each group representing a layer of MgO
on the left side of the nanogap (labeled as L1, L2, L3,
and L4). We illustrate the variation of local VDOS around the MgO-MgO nanogap along the transport direction as presented in Fig.~\ref{fig:vdosl}. Notably, the local VDOS in the L1 region differ greatly from those in the bulk MgO crystal. The change in surface phonon properties is much greater than in the case of contact, possibly due to the lack of constraints on atomic motion near the vacuum. Furthermore, the local VDOS is highly sensitive to the arrangement of atoms at the surface, which greatly influences heat transport, as discussed in the main text. Subsequently, an analysis was conducted of the transition of local VDOS from bulk to surface. As anticipated, the local VDOS of the material shifts closer to that of the bulk material with increasing distance from the nanogap interface.
	
\end{document}